\documentclass[sigconf]{acmart}
\usepackage{algorithm}
\usepackage{algorithmic}
\usepackage{amsfonts}
\usepackage{amsmath}
\usepackage{booktabs}
\usepackage{array}
\usepackage{multirow}
\usepackage{tabularx}
\usepackage{xcolor}
\usepackage{color, colortbl}
\usepackage{adjustbox}
\usepackage{arydshln}
\usepackage{newfloat}
\usepackage{listings}
\copyrightyear{2026}
\acmYear{2026}
\setcopyright{cc}
\setcctype{by}
\acmConference[SIGIR '26] {Proceedings of the 49th International ACM SIGIR Conference on Research and Development in Information Retrieval}{July 20--24, 2026}{Melbourne, VIC, Australia.}
\acmBooktitle{Proceedings of the 49th International ACM SIGIR Conference on Research and Development in Information Retrieval (SIGIR '26), July 20--24, 2026, Melbourne, VIC, Australia}
\acmISBN{979-8-4007-2599-9/2026/07}
\acmDOI{10.1145/3805712.3809977}

\begin{document}

\title{Structural and Disentangled Adaptation of Large Vision Language Models for Multimodal Recommendation}

\author{Zhongtao Rao}
\authornote{Equal contribution.}
\email{zrao690@connect.hkust-gz.edu.cn}
\affiliation{
  \institution{The Hong Kong University of Science and Technology (Guangzhou)}
  \country{Guangzhou}
  \country{China}
}

\author{Peilin Zhou}
\authornotemark[1]   
\email{zhoupalin@gmail.com}
\affiliation{
  \institution{New York University Abu Dhabi}
    \city{Abu Dhabi}
  \country{United Arab Emirates}
}

\author{Dading Chong}
\email{1601213984@pku.edu.cn}
\affiliation{
  \institution{Peking University}
  \city{Beijing}
  \country{China}
}

\author{Zhiwei Chen}
\email{zhiweic@hkust-gz.edu.cn}
\affiliation{
  \institution{The Hong Kong University of Science and Technology (Guangzhou)}
  \country{Guangzhou}
  \country{China}
}
\author{Shoujin Wang}
\email{shoujin.wang@uts.edu.au}
\affiliation{
  \institution{University of Technology Sydney}
  \city{Sydney}
  \country{Australia}
}

\author{Nan Tang}
\email{nantang@hkust-gz.edu.cn}
\affiliation{
  \institution{The Hong Kong University of Science and Technology (Guangzhou)}
  \city{Guangzhou}
  \country{China}
}
\authornote{Corresponding author.}  

\renewcommand{\shortauthors}{Zhongtao et al.}

\begin{abstract}
Multimodal recommendation enhances accuracy by leveraging visual and textual signals, 
and its success largely depends on learning high-quality cross-modal representations. 
Recent advances in Large Vision-Language Models (LVLMs) offer unified multimodal representation learning, making them a promising backbone. 
However, applying LVLMs to recommendation remains challenging due to (\textit{i}) \textit{representation misalignment}, where gaps between specific domain data and general pre-training lead to unaligned embedding spaces, and (\textit{ii}) \textit{gradient conflicts} during fine-tuning, where shared adapters cause interference and a lack of discriminative power.
To address this, we propose \textbf{SDA}, a lightweight framework for \underline{\textbf{S}}tructural and \underline{\textbf{D}}isentangled \underline{\textbf{A}}daptation, which integrates two components: \emph{Cross-Modal Structural Alignment} (CMSA) and \emph{Modality-Disentangled Adaptation} (MoDA). 
CMSA aligns embeddings using intra-modal structures as a soft teacher, while MoDA mitigates gradient conflicts via expertized, gated low-rank paths to disentangle gradient flows. 
Experiments on three public Amazon datasets show \textbf{SDA} integrates seamlessly with existing multimodal and sequential recommenders, yielding average gains of \textbf{6.15\%} in Hit@10 and \textbf{8.64\%} in NDCG@10. 
It also achieves up to \textbf{12.83\%} and \textbf{18.70\%} gains on long-tail items. 
Our code and full experimental results are available at \url{https://github.com/RaoZhongtao/SDA}.

\end{abstract}




\maketitle
\section{Introduction}\label{sec:intro}
\begin{figure}[t]
\centering
\includegraphics[width=1\columnwidth]{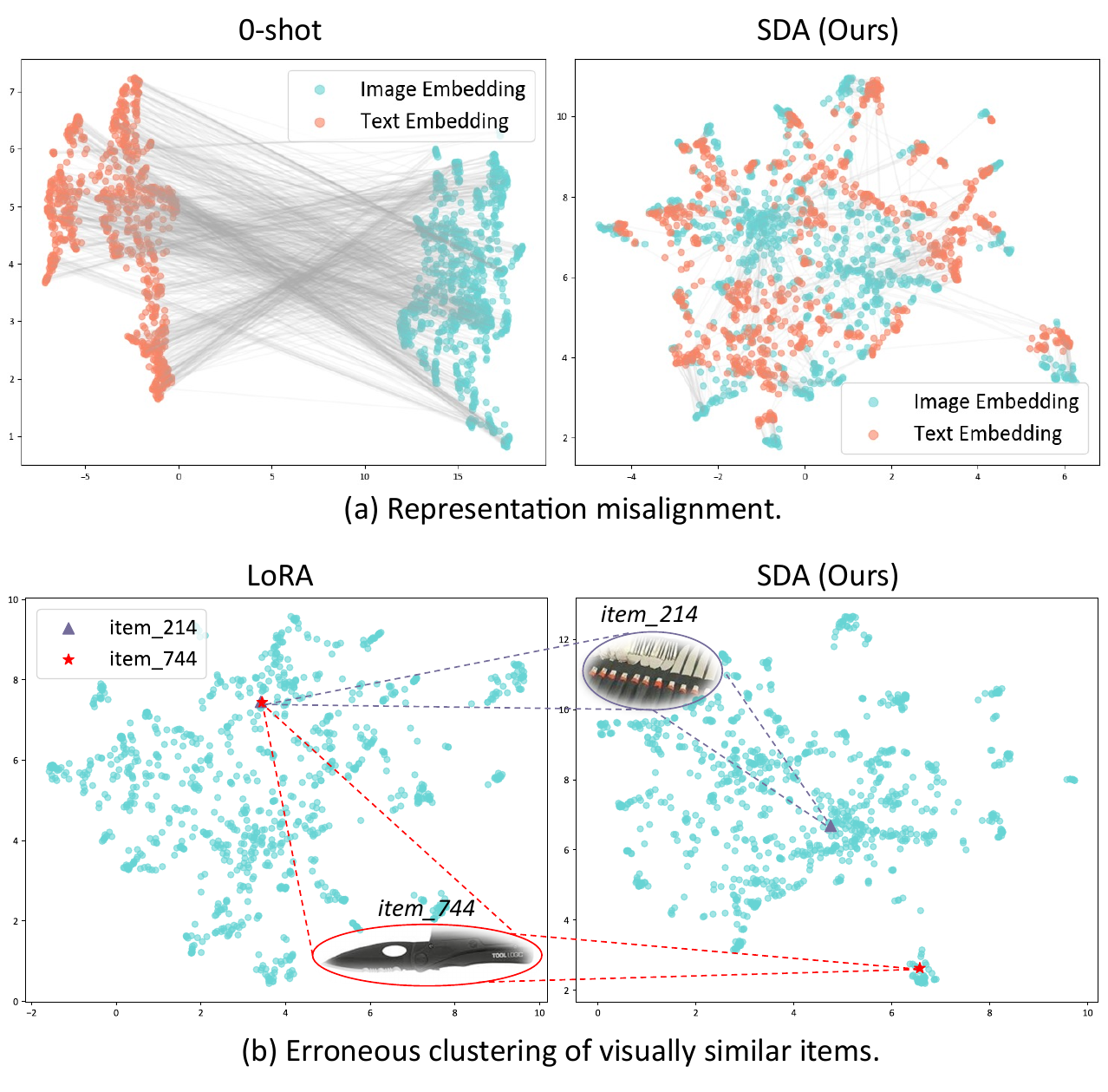}
\caption{Illustration of two key challenges when applying LVLMs to recommendation.}
\label{Figure:motivation}
\end{figure}

Multimodal recommendation enhances personalized systems by leveraging visual and textual content to capture item characteristics and user preferences~\cite{liu2024multimodal,zhou2025large}. Early approaches fused features from separate encoders (e.g., CNNs, Transformers) or used dual-encoders like CLIP~\cite{radford2021learning}.
Recent advances in Large Vision--Language Models (LVLMs), such as GPT-4o and Qwen-VL~\cite{Qwen2.5-VL}, offer a more powerful alternative. Compared to earlier models, their massive scale, unified architectures, and deep cross-modal attention allow them to excel at modeling fine-grained visual--textual relationships and transferring rich world knowledge~\cite{yin2024survey}. This makes them attractive as multimodal feature extractors for ID-based recommenders.

However, applying LVLMs in this way is non-trivial, facing a two-fold challenge. The first is \textit{representation misalignment} (C1). This arises because the domain-specific nature of item images and texts (e.g., product photos, short titles) differs significantly from domain-free pre-training data. As shown in Figure~\ref{Figure:motivation}(a), this results in unaligned image and text embedding spaces, creating a large semantic gap for the same item. Such misalignment necessitates fine-tuning, yet this adaptation process is hindered by a second challenge: \textit{gradient conflicts} (C2). Parameter-efficient tuning methods (e.g., LoRA~\cite{hu2022lora}) typically share low-rank adapters across modalities, causing interference between visual and textual gradients~\cite{yu2023multi}. As illustrated in Figure~\ref{Figure:motivation}(b), this conflict results in a lack of discriminative power, causing items that are visually similar but functionally different to be mistakenly clustered together.

To address these challenges, we propose \textbf{SDA}, a lightweight framework adapting LVLMs for recommendation via \textbf{\underline{S}}tructural and \textbf{\underline{D}}isentangled \textbf{\underline{A}}daptation. Our framework systematically tackles both issues successively. To tackle the initial misalignment (C1), we design \emph{Cross-Modal Structural Alignment} (CMSA). CMSA leverages meaningful, preserved intra-modal relations (which remain intact even when cross-modal alignment is lost) as a soft teacher to align the disparate text-image embedding spaces at the distribution level. To resolve the subsequent gradient conflicts (C2), we design \emph{Modality-Disentangled Adaptation} (MoDA). MoDA targets the root cause of the conflict, which is the shared adapter, by replacing it with expertized, gated low-rank paths to effectively disentangle visual and textual gradient flows. We instantiate \textbf{SDA} using Qwen-VL as the LVLM backbone and integrate it with two multimodal and two sequential recommenders. Experiments on three public datasets show that it consistently improves performance, achieving average gains of \textbf{6.15\%} in Hit@10 and \textbf{8.64\%} in NDCG@10, and up to \textbf{12.83\%} and \textbf{18.70\%} relative gains on long-tail items, while incurring only minimal inference overhead.

\begin{figure*}[t]
\centering
\includegraphics[width=0.95\textwidth]{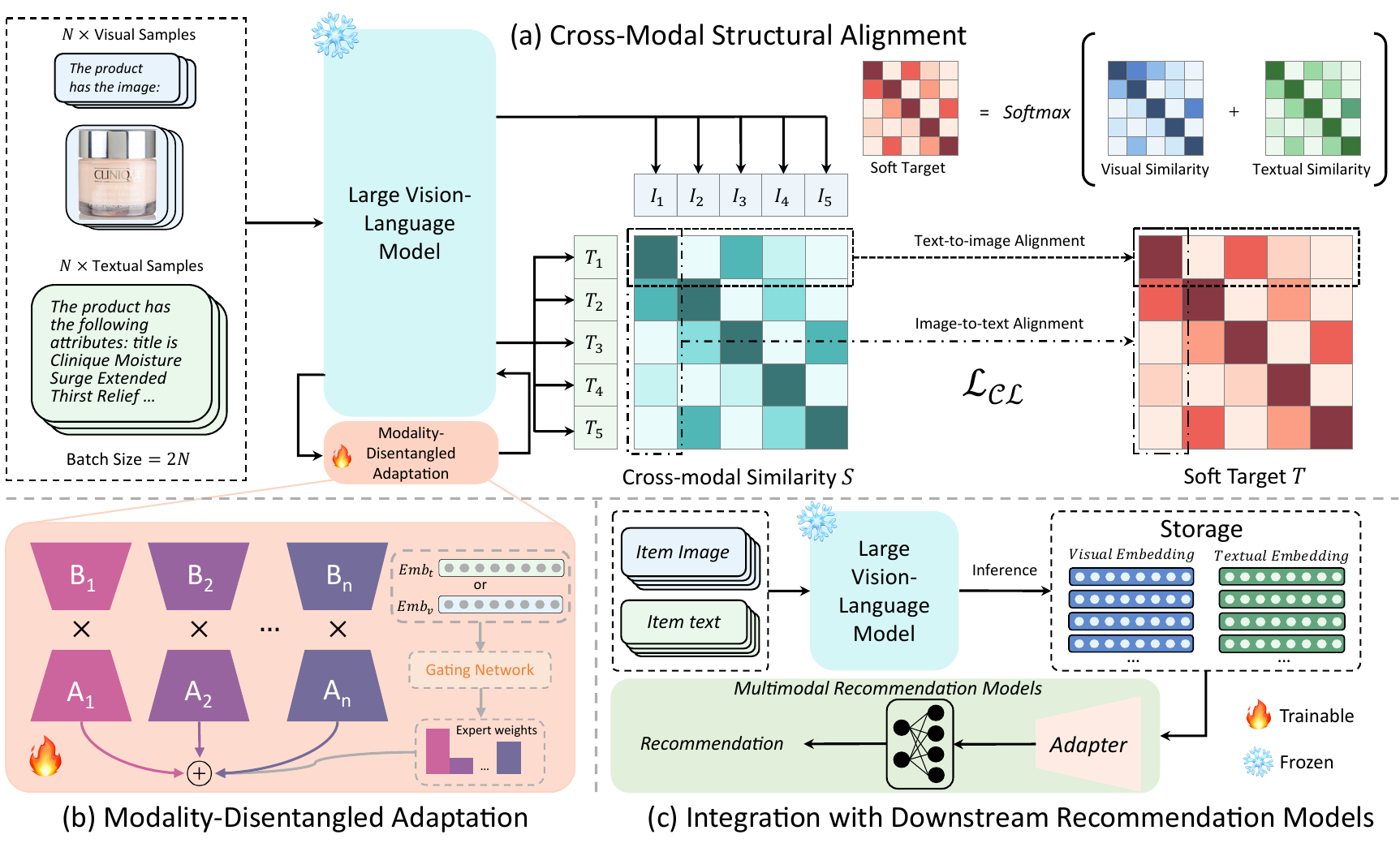}
\caption{Overview of the proposed SDA framework.}
\label{Figure:SDA}
\end{figure*}

\section{Methodology}
As shown in Figure~\ref{Figure:SDA}, SDA contains two components: \emph{Cross-Modal Structural Alignment} (CMSA) and \emph{Modality-Disentangled Adaptation} (MoDA). CMSA adapts LVLM representations to recommendation data via a structure-aware contrastive objective, and MoDA performs parameter-efficient, modality-specific tuning. The adapted LVLM then serves as a multimodal feature extractor for downstream recommenders.

\subsection{Cross-Modal Structural Alignment (CMSA)}

CMSA adapts LVLM representations by aligning text--image pairs while enforcing that items form a shared, fine-grained cross-modal neighborhood structure. 
For each item, we construct simple text and image prompts: the textual prompt concatenates basic fields (e.g., title, category, brand) with a short instruction "\textit{The product has the following attributes:}", while the visual prompt introduces the product image with "\textit{The product has the image:}". 
Both prompts are fed into a pretrained LVLM, and we apply mean pooling over the final hidden states followed by a projection to obtain $d_m$-dimensional embeddings $\mathbf{e}_i^t, \mathbf{e}_i^v \in \mathbb{R}^{d_m}$ for the textual and visual modalities of item $i$, respectively. 

Given a batch of $N$ items, we first compute the cross-modal similarity matrix:
\begin{equation}
    \mathbf{S}_{ij} = \frac{\mathbf{e}_i^t \cdot \mathbf{e}_j^v}{\tau}, \quad i,j \in \{1,\ldots,N\},
\end{equation}
where $\tau$ is a temperature hyperparameter. To construct the structural teacher, we build a soft target distribution $\mathbf{T} \in \mathbb{R}^{N \times N}$ from averaged intra-modal similarities:
\begin{equation}
    \mathbf{T}_{ij} = \text{Softmax}_j\!\left( \frac{\tau}{2} \big( \mathbf{e}_i^t \cdot \mathbf{e}_j^t + \mathbf{e}_i^v \cdot \mathbf{e}_j^v \big) \right),
\end{equation}
where $\text{Softmax}_j(\cdot)$ denotes row-wise normalization over $j$. For each anchor item $i$, the row $\mathbf{T}_{i,:}$ encodes its neighborhood structure over all other items by combining text- and image-based similarities.

CMSA then performs distribution-level alignment by matching the cross-modal similarity distributions induced by $\mathbf{S}$ to this structural teacher. Let $\mathbf{P}_{ij} = \text{Softmax}_j(\mathbf{S}_{ij})$ and $\mathbf{P}^{\top}$ be the analogous distribution over $\mathbf{S}^\top$. The structure-aware contrastive loss is:
\begin{equation}
    \mathcal{L}_{\text{CL}} = \frac{1}{2N} \sum_{i=1}^N \Big( \mathrm{KL}(\mathbf{T}_{i,:} \Vert \mathbf{P}_{i,:}) + \mathrm{KL}(\mathbf{T}_{:,i} \Vert \mathbf{P}^{\top}_{:,i}) \Big),
\end{equation}
where $\mathrm{KL}(\cdot\Vert\cdot)$ denotes Kullback--Leibler divergence. By aligning entire cross-modal similarity distributions with this structure-aware soft target, CMSA transfers intra-modal relational knowledge into the joint space and yields more discriminative, modality-invariant item embeddings.

\subsection{Modality-Disentangled Adaptation (MoDA)}

MoDA is a modality-aware low-rank adaptation framework that preserves the efficiency of LoRA~\cite{hu2022lora} while explicitly routing gradients through modality-specific expert combinations. Given a target weight matrix $\mathbf{W}_0 \in \mathbb{R}^{d_{\text{in}} \times d_{\text{out}}}$ in the frozen LVLM, standard LoRA introduces a low-rank update:
\begin{equation}
    \Delta \mathbf{W} = BA,
\end{equation}
where $B \in \mathbb{R}^{d_{\text{in}} \times r}$ and $A \in \mathbb{R}^{r \times d_{\text{out}}}$ with $r \ll \min(d_{\text{in}}, d_{\text{out}})$. In MoDA, we factorize this low-rank space into $N_e$ experts and let each modality softly combine them:
\begin{equation}
    \Delta \mathbf{W} = \sum_{i=1}^{N_e} \omega_i B_i A_i,
\end{equation}
where $B_i \in \mathbb{R}^{d_{\text{in}} \times r_e}$ and $A_i \in \mathbb{R}^{r_e \times d_{\text{out}}}$ are the parameters of expert $i$, $r_e = r / N_e$, and $\boldsymbol{\omega} = (\omega_1, \ldots, \omega_{N_e}) \in \mathbb{R}^{N_e}$ are expert weights. 

For each modality $m$ (e.g., vision or text), we introduce a trainable modality embedding $\text{Emb}_m$, which is fed into a small gating network:
\begin{equation}
    \boldsymbol{\omega}^{m} = \text{Gate}(\text{Emb}_m) \in \mathbb{R}^{N_e}.
\end{equation}
The resulting forward computation for input $x$ from modality $m$ is:
\begin{equation}
    h = \mathbf{W}_0 x + \sum_{i=1}^{N_e} \omega_i^{m} B_i A_i x.
\end{equation}
Thus, different modalities share the same pool of experts but emphasize different subsets or combinations through $\boldsymbol{\omega}^{m}$, without maintaining fully separate adapters. This modality-aware routing disentangles visual and textual gradient flows in the low-rank subspace, substantially reducing cross-modal interference while keeping the overall rank $r$ (and hence parameter count) comparable to standard LoRA.

\subsection{Integration and Model Complexity}

We adopt a two-stage pipeline: we first insert MoDA into the LVLM and optimize only its parameters with respect to $\mathcal{L}_{\text{CL}}$ on item data to obtain adapted text and image embeddings $\mathbf{e}_i^t$ and $\mathbf{e}_i^v$, and then freeze the LVLM, precompute these embeddings offline, and feed them into existing multimodal or sequential recommenders (e.g., via concatenation or other fusion in an adapter module). Since SDA only adds a small set of low-rank experts on top of the frozen LVLM (CMSA itself introduces no extra trainable parameters) and all item embeddings are precomputed, the number of additional trainable parameters is a small fraction of the LVLM backbone and the serving-time complexity of downstream recommenders remains essentially unchanged.

\section{Experiments}

\subsection{Experimental Settings}

\begin{table*}[h]
\caption{Performance comparison of SDA features against Base, CLIP, and QwenVL (vanilla Qwen2.5-VL 7B) on three Amazon datasets. Best results are in boldface, second best are underlined. $^{*}$ denotes $p<0.05$ from paired t-test.}
\centering
\footnotesize
\begin{tabularx}{\textwidth}{p{1.3cm} p{1.0cm} *{12}{>{\centering\arraybackslash}X}}
\toprule
\multirow{3}{*}{\textbf{Model}} & \multirow{3}{*}{\textbf{Extractor}} 
& \multicolumn{4}{c}{\textbf{Beauty}} 
& \multicolumn{4}{c}{\textbf{Sports}} 
& \multicolumn{4}{c}{\textbf{Toys}} \\
\cmidrule(lr){3-6} \cmidrule(lr){7-10} \cmidrule(lr){11-14}
& & \multicolumn{2}{c}{Overall} & \multicolumn{2}{c}{Tail}
& \multicolumn{2}{c}{Overall} & \multicolumn{2}{c}{Tail}
& \multicolumn{2}{c}{Overall} & \multicolumn{2}{c}{Tail} \\
\cmidrule(lr){3-4} \cmidrule(lr){5-6}
\cmidrule(lr){7-8} \cmidrule(lr){9-10}
\cmidrule(lr){11-12} \cmidrule(lr){13-14}
& & H@10 & N@10 & H@10 & N@10 & H@10 & N@10 & H@10 & N@10 & H@10 & N@10 & H@10 & N@10 \\
\midrule

\multirow{3}{*}{\textbf{SLMRec}} 
& Base 
& 0.4258 & 0.2852 & 0.2003 & 0.1020 & 0.4632 & 0.2962 & \underline{0.2926} & \underline{0.1429} & 0.4617 & \underline{0.3115} & 0.3315 & \underline{0.1890} \\
& CLIP 
& \underline{0.4363} & \underline{0.2887} & \underline{0.2127} & \underline{0.1044} & \underline{0.4710} & \underline{0.3024} & 0.2595 & 0.1250 & \underline{0.4687} & 0.3107 & \underline{0.3319} & 0.1830 \\
& QwenVL 
& 0.4187 & 0.2652 & 0.1856 & 0.0917 & 0.4337 & 0.2636 & 0.1833 & 0.0848 & 0.4502 & 0.2900 & 0.2727 & 0.1471 \\
 & \textbf{SDA}
& \textbf{0.4805$^{*}$} & \textbf{0.3295$^{*}$} & \textbf{0.3191$^{*}$} & \textbf{0.1766$^{*}$}
& \textbf{0.5115$^{*}$} & \textbf{0.3367$^{*}$} & \textbf{0.3545$^{*}$} & \textbf{0.1880$^{*}$}
& \textbf{0.5048$^{*}$} & \textbf{0.3453$^{*}$} & \textbf{0.3988$^{*}$} & \textbf{0.2357$^{*}$} \\
\midrule

\multirow{3}{*}{\textbf{VBPR}}
& Base 
& 0.4572 & 0.3112 & 0.0804 & 0.0419 & 0.4969 & 0.3307 & 0.0839 & 0.0399 & 0.4536 & 0.3061 & 0.1252 & 0.0751 \\
& CLIP 
& \underline{0.5093} & \underline{0.3506} & 0.3625 & 0.2251 & \underline{0.5416} & \underline{0.3496} & 0.3890 & 0.2226 & \underline{0.5329} & \underline{0.3707} & 0.4501 & 0.3052 \\
& QwenVL 
& 0.4379 & 0.2877 & \underline{0.3824} & \underline{0.2437} & 0.4480 & 0.2697 & \underline{0.3921} & \underline{0.2361} & 0.4692 & 0.3167 & \underline{0.4502} & \underline{0.3060}  \\
 & \textbf{SDA} 
& \textbf{0.5375$^{*}$} & \textbf{0.3745$^{*}$} & \textbf{0.3981$^{*}$} & \textbf{0.2570$^{*}$} 
& \textbf{0.5731$^{*}$} & \textbf{0.3781$^{*}$} & \textbf{0.4306$^{*}$} & \textbf{0.2598$^{*}$}
& \textbf{0.5556$^{*}$} & \textbf{0.3915$^{*}$}  & \textbf{0.4761$^{*}$} & \textbf{0.3308$^{*}$} \\
\midrule

\multirow{4}{*}{\textbf{SASRec}} 
& Base 
& 0.4141 & 0.2760 & 0.0462 & 0.0229 & 0.4441 & 0.2882 & 0.0462 & 0.0216 & 0.4126 & 0.2758 & 0.0782 & 0.0452 \\
& CLIP 
& 0.5331 & 0.3524 & \underline{0.4336} & \underline{0.2719} & 0.5760 & 0.3539 & \underline{0.4906} & \underline{0.2927} & 0.5800 & 0.3998 & \underline{0.5292} & \underline{0.3659} \\
& QwenVL 
& \underline{0.5569} & \underline{0.3675} & 0.4277 & 0.2590 & \underline{0.5952} & \underline{0.3708} & 0.4755 & 0.2860 & \underline{0.6106} & \underline{0.4225} & 0.5172 & 0.3541 \\
 & \textbf{SDA}
& \textbf{0.5752$^{*}$} & \textbf{0.3951$^{*}$} & \textbf{0.4692$^{*}$} & \textbf{0.3118$^{*}$}
& \textbf{0.6192$^{*}$} & \textbf{0.3964$^{*}$} & \textbf{0.5209$^{*}$} & \textbf{0.3284$^{*}$}
& \textbf{0.6125$^{*}$} & \textbf{0.4333$^{*}$}  & \textbf{0.5522$^{*}$} & \textbf{0.3914$^{*}$} \\
\midrule

\multirow{4}{*}{\textbf{BERT4Rec}}
& Base  
& 0.3955 & 0.2349 & 0.0031 & 0.0011 & 0.3877 & 0.2246 & 0.0017 & 0.0006 & 0.3554 & 0.2032 & 0.0045 & 0.0017 \\
& CLIP 
& \underline{0.5380} & \underline{0.3468} & \textbf{0.4071$^{*}$} & \textbf{0.2435$^{*}$} & \underline{0.5684} & \underline{0.3350} & \textbf{0.4425$^{*}$} & \textbf{0.2439$^{*}$} & \underline{0.5843} & \underline{0.3896} & \underline{0.5084} & \underline{0.3330} \\
& QwenVL 
& 0.4913 & 0.3027 & 0.3236 & 0.1737 & 0.4952 & 0.2793 & 0.3160 & 0.1609 & 0.5352 & 0.3361 & 0.3976 & 0.2360 \\
 &  \textbf{SDA}
& \textbf{0.5535$^{*}$} & \textbf{0.3521$^{*}$} & \underline{0.4070} & \underline{0.2329} 
& \textbf{0.5847$^{*}$} & \textbf{0.3515$^{*}$} & \underline{0.4344} & \underline{0.2413}
& \textbf{0.6138$^{*}$}  & \textbf{0.4190$^{*}$}  & \textbf{0.5288$^{*}$}  & \textbf{0.3609$^{*}$} \\
\bottomrule
\end{tabularx}
\label{tab:overall_performance}
\end{table*}

\begin{table}
  \begin{tabular}{ccccc}
    \toprule
    Dataset & \#Users & \#Items & \#Interactions & Sparsity \\
    \midrule
    Beauty & 50,498 & 57,019 & 391,956 & 99.986\% \\
    Sports & 80,097 & 83,006 & 582,372 & 99.991\% \\
    Toys & 53,983 & 68,556 & 404,955 & 99.989\% \\
  \bottomrule
\end{tabular}
\caption{Statistics of the datasets.}
\label{tab:statistics}
\end{table}

\subsubsection{Datasets and Evaluation Metrics} 
We conduct experiments on three categories from the widely used Amazon Reviews dataset~\cite{mcauley2015image}: \textit{Beauty}, \textit{Sports}, and \textit{Toys}. Detailed statistics for these datasets is shown as Table~\ref{tab:statistics}.
We follow the standard leave-one-out protocol~\cite{kang2018self,zhou2023equivariant,zhou2023attention}.
During evaluation, each ground-truth item is ranked among 100 negative items randomly sampled from those the user has not interacted with. We report two widely-used metrics: Hit Rate@10 (H@10) and Normalized Discounted Cumulative Gain@10 (N@10).  To further assess model performance on long-tail items, we define items with fewer than four interactions as long-tail items. Users whose ground-truth item belongs to this category are grouped into the long-tail user set. We evaluate the model on this subset and report the corresponding metrics as \textit{Tail} H@10 and \textit{Tail} N@10.

\subsubsection{Baselines}
We integrate SDA framework into four backbone recommenders: two multimodal (SLMRec~\cite{Tao2023SLMRec}, VBPR~\cite{he2016vbpr}) and two sequential (SASRec~\cite{kang2018self}, BERT4Rec~\cite{Sun2019bert4rec}). As shown in Table~\ref{tab:overall_performance}, we compare SDA-generated features against three baseline feature settings: \textit{Base} (the backbones' original features, e.g., learnable ID embeddings for SASRec), \textit{CLIP} (features from a pre-trained CLIP model), and \textit{QwenVL} (zero-shot features from the vanilla Qwen-VL model). Both \textit{QwenVL} and our \textbf{SDA} framework utilize \textit{Qwen2.5-VL 7B Instruct}~\cite{Qwen2.5-VL} as the LVLM backbone. 

\subsubsection{Implementation details}
All multimodal baseline implementations and evaluations are conducted using MMRec~\cite{zhou2023mmrec}, a unified open-source framework for multimodal recommendation. For SASRec and BERT4Rec, we adopt the implementations provided by LLMEmb~\cite{liu2025llmemb}. In addition, we employ the RAT method from LLMEmb to effectively adapt SDA-generated embeddings for compatibility with these sequential recommenders. All experiments are conducted on a server with Intel Xeon Platinum 8358P CPUs and NVIDIA A800 80GB GPUs.
For fair comparison, the embedding dimension of all models is set to 128. Following prior work, all baseline multimodal models adopt \textit{all-MiniLM-L6-v2} from Sentence-Transformers for textual features, and the visual features provided in the Amazon Reviews dataset~\cite{mcauley2015image}, which are extracted using \textit{AlexNet}. The CLIP model we used is \textit{clip-vit-large-patch14}. For models implemented within MMRec, we use a batch size of 2048, with early stopping patience of 20 and a maximum of 1000 epochs. For each model, we conduct exhaustive grid search over the recommended hyperparameter configurations to ensure optimal performance.
For SDA, we fine-tune the model for at most 4000 steps to avoid overfitting, updating all linear layers while keeping the vision encoder (ViT) frozen. We set the number of experts in SDA to 8, which yields the best performance, and use a training batch size of 32.

\subsection{Overall Performance}
Table~\ref{tab:overall_performance} presents our experimental results, from which we derive three key observations. 
\textbf{First, SDA achieves the strongest performance on overall metrics.} 
Across all four backbones (SLMRec, VBPR, SASRec, and BERT4Rec), SDA consistently obtains the best "Overall" H@10 and N@10 results, highlighting its ability to generate robust representations for general user preferences.
\textbf{Second, SDA delivers substantial gains in long-tail recommendation.}
Across all backbones and datasets, SDA consistently outperforms both the Base and QwenVL baselines on tail metrics. 
Furthermore, in the vast majority of scenarios, SDA significantly surpasses the strong general-purpose CLIP features. 
This demonstrates SDA's effectiveness in mitigating data sparsity and improving representation quality for cold-start or infrequent items, even when integrated into strong sequential models like SASRec and BERT4Rec.
\textbf{Lastly, vanilla LVLMs are not a "plug-and-play" solution, confirming the need for adaptation.}
The performance of zero-shot QwenVL is unstable; it is outperformed by the simpler CLIP model in most cases and sometimes performs worse than the Base model (e.g., 0.4337 vs. 0.4632 H@10 for SLMRec on Sports). 
This strongly validates our \textit{representation misalignment} (C1) diagnosis: the "domain gap" between the LVLM's general pre-training and recommendation semantics leads to negative transfer. 
SDA systematically addresses this gap, unlocking the LVLM's potential and transforming it into the best-performing model for overall recommendation.
\subsection{In-depth Analysis}
\subsubsection{Ablation Study} Table~\ref{tab:ablation} details the ablation analysis on \textit{Beauty}. We first examine the impact of individual components. Removing CMSA results in a significant 6.34\% average drop. Crucially, replacing our structure-aware soft target with standard InfoNCE (\textit{w/o Soft Target}) leads to an even steeper 8.41\% decline, confirming that preserving fine-grained intra-modal structure is superior to simple alignment. Regarding MoDA, replacing it with standard LoRA (\textit{w/o MoDA}) reduces performance by 2.65\%, validating the efficacy of our expertized gating strategy in mitigating gradient conflicts. Finally, when both components are removed, performance suffers a catastrophic 29.08\% drop, effectively reverting to the unadapted baseline and demonstrating their combined indispensability. 

\begin{table}[t]
\caption{Ablation study on the \textit{Beauty} dataset. $\Delta$ denotes the average relative performance drop of H@10 and N@10 compared to the full model.}
\centering
\resizebox{\columnwidth}{!}{
\begin{tabular}{lcccccc}
\toprule
\multirow{2}{*}{Method} & \multicolumn{3}{c}{Overall} & \multicolumn{3}{c}{Tail} \\
\cmidrule(lr){2-4} \cmidrule(lr){5-7}
 & H@10 & N@10 & \(\Delta\) & H@10 & N@10 & \(\Delta\) \\
\midrule
\textit{\textbf{SDA}} & \textbf{0.5752} & \textbf{0.3951} & 0.00\% & \textbf{0.4692} & \textbf{0.3118} & 0.00\% \\
\textit{w/o CMSA} & 0.5517 & 0.3612 & -6.34\% & 0.4266 & 0.2622 & -12.50\% \\
\textit{w/o MoDA} & 0.5604 & 0.3843 & -2.65\% & 0.4496 & 0.2969 & -4.48\% \\
w/o CMSA and MoDA & 0.4141 & 0.2760 & -29.08\% & 0.0229 & 0.0462 & -90.15\% \\
\textit{w/o Soft Target} & 0.5485 & 0.3470 & -8.41\% & 0.4291 & 0.2512 & -14.00\% \\
\bottomrule
\end{tabular}
}
\label{tab:ablation}
\end{table}

\subsubsection{Modality Contribution Analysis.}
\begin{figure}[t]
\centering
\includegraphics[width=0.9\columnwidth]{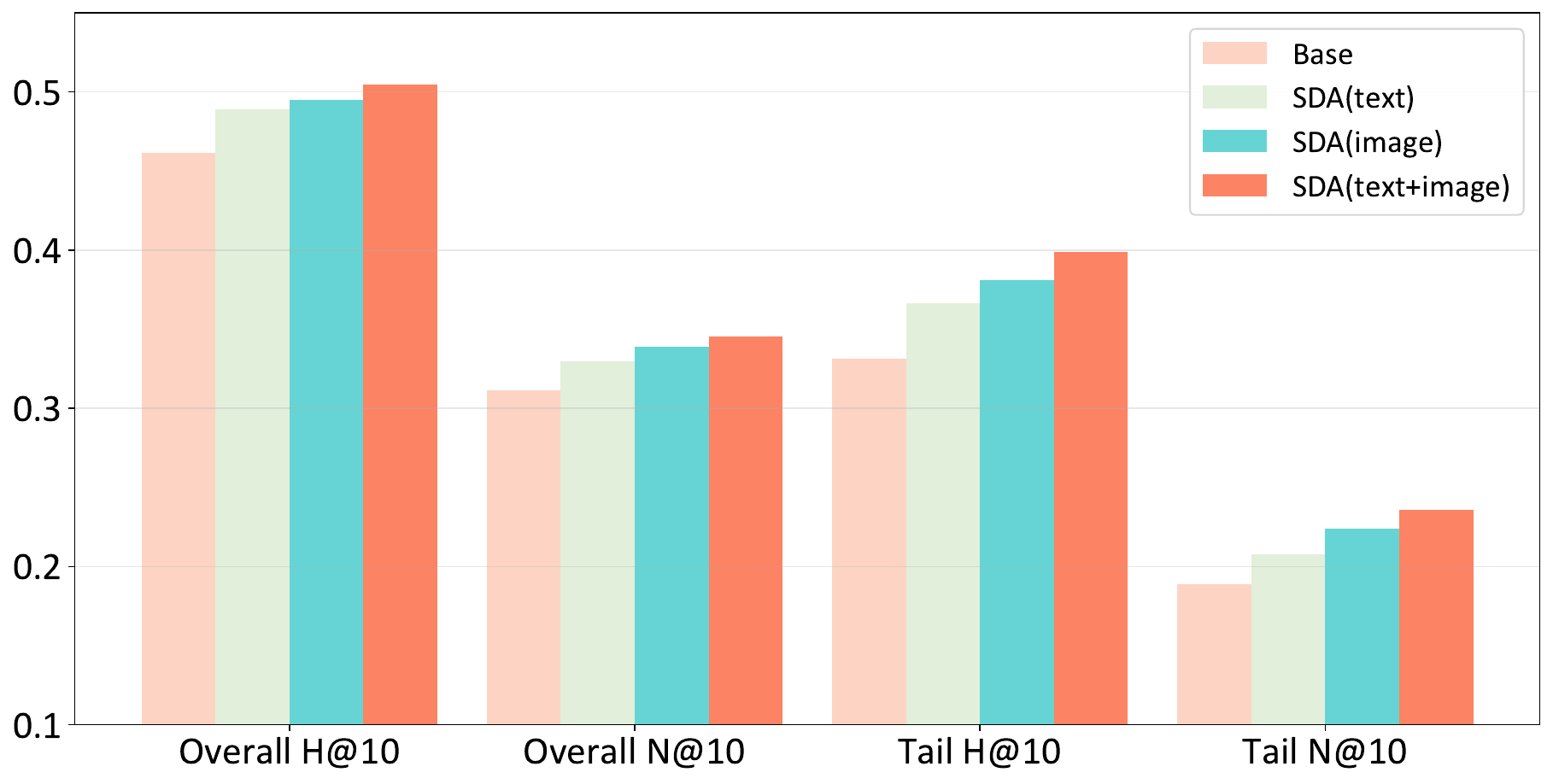}
\caption{Impact of individual and combined modalities on the \textit{Toys} dataset with SLMRec as backbone.}
\label{tab:modality_compare}
\end{figure}

We evaluate the contribution of SDA-generated features on the \textit{Toys} dataset using SLMRec, as shown in Figure~\ref{tab:modality_compare}. Individually, both SDA-adapted textual and visual features outperform the original SLMRec baseline, with visual features providing larger gains. Crucially, the combination of both modalities yields the highest performance across all metrics. This demonstrates that SDA effectively captures complementary information, resulting in a synergistic improvement when integrating both visual and textual signals.

\subsubsection{Gradient Conflict}

We investigate gradient conflicts by analyzing the cosine similarity of modality-specific gradients in the last decoder layer of Qwen2.5-VL. To isolate visual and textual gradients from the scalar contrastive loss, we employ \texttt{torch.detach()} to block alternative modalities during backpropagation. Focusing on the LoRA $B$ matrices in the \textit{q\_proj} and \textit{k\_proj} layers (using the first expert for MoDA), we observe a striking contrast: standard LoRA exhibits negative similarity (–0.0955 for \textit{q\_proj}, –0.0705 for \textit{k\_proj}), confirming the presence of severe cross-modal interference. In contrast, MoDA yields strong positive similarity (0.4422 and 0.7096), demonstrating that our disentangled adaptation successfully fosters compatible and aligned updates.

\begin{figure}[t]
\centering
\includegraphics[width=0.45\textwidth]{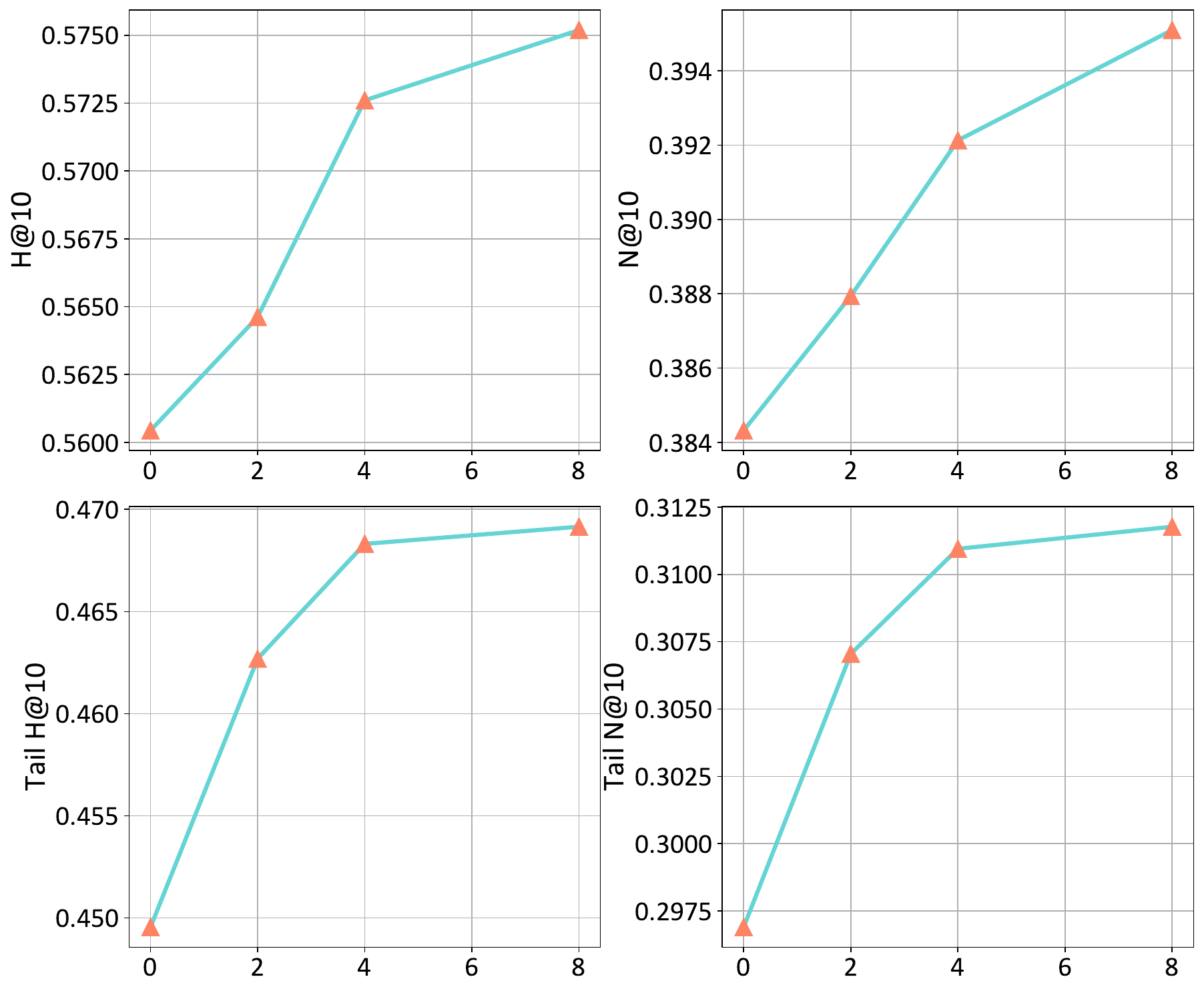}
\caption{Performance on the \textit{Beauty} dataset with different $N_e$. $N_e = 0$ means using standard LoRA.}
\label{Figure:num_experts}
\end{figure}

\subsubsection{Hyperparameter Analysis}
We conducted experiments varying the number of experts (denoted as $N_e$) on the MoDA. The LoRA rank is fixed at 8, and $N_e$ is set to 2, 4, or 8. For comparison, we also included a baseline where $N_e = 0$, which corresponds to standard LoRA without expert specialization. The results on \textit{Beauty} dataset are shown in Figure~\ref{Figure:num_experts}, which demonstrate that increasing the number of experts consistently improves model performance. 
This aligns with our expectation that a larger number of experts enables the model to learn more expressive and modality-specific representations.

\section{Conclusion and Future Work}
In this paper, we propose \textbf{SDA}, a lightweight framework designed to adapt LVLMs for multimodal recommendation. By integrating Cross-Modal Structural Alignment (CMSA) and Modality Disentangled Adaptation (MoDA), SDA effectively resolves the twin challenges of representation misalignment and gradient conflicts. Extensive experiments demonstrate that SDA generates robust, transferable representations that consistently enhance various downstream recommenders. Notably, it yields significant gains in long-tail scenarios, proving to be a versatile and effective solution for unlocking the potential of LVLMs in recommendation tasks.

Despite its effectiveness, this work has several limitations that also point to promising directions for future research. 
First, our experiments are conducted on three categories from the Amazon Reviews dataset, which may not fully reflect the diversity of multimodal recommendation scenarios. 
Extending the evaluation to other domains, such as short-video or news recommendation, would provide a more comprehensive understanding of the applicability of SDA.
Second, while the performance gap between vanilla LVLM features and SDA-adapted representations provides empirical evidence of representation misalignment, a more direct and systematic quantification of cross-modal alignment remains an open problem. 
Developing explicit metrics or diagnostic tools to characterize modality gaps could further enhance the interpretability of adaptation methods.




\begin{acks}
This work is supported by Guangdong provincial project (Project No. 2023CX10X008). It is also 
benefited greatly from the generous support, resources, and funding provided by the Red Bird MPhil Program at the Hong Kong University of Science and Technology (Guangzhou), which were instrumental in the successful completion of this research.
\end{acks}

\bibliographystyle{ACM-Reference-Format}
\balance
\bibliography{sample-base}

\appendix

\end{document}